\documentclass[12pt]{article}

\usepackage{float}
\usepackage{graphics, graphicx}
\usepackage{amsfonts}
\usepackage{natbib}

\usepackage[margin=10pt,font=small,labelfont=bf,labelsep=period]{caption}

\usepackage{amsmath,amsbsy,amssymb,amsthm,latexsym, tabularx}
\usepackage{multirow,multicol,subfigure,booktabs,url}

\newcommand{\bt}{\begin{itemize}}
\newcommand{\et}{\end{itemize}}
\newcommand{\ben}{\begin{enumerate}}
\newcommand{\een}{\end{enumerate}}
\newcommand{\beq}{\begin{equation}}
\newcommand{\eeq}{\end{equation}}
\newcommand{\beqn}{\begin{eqnarray}}
\newcommand{\eeqn}{\end{eqnarray}}
\newcommand{\bea}{\begin{eqnarray*}}
\newcommand{\eea}{\end{eqnarray*}}
\newcommand{\bs}{\boldsymbol}
\newcommand{\argmax}{\operatornamewithlimits{argmax}}

\usepackage{algorithm}
\usepackage[noend]{algpseudocode}

\usepackage{setspace}
\let\Algorithm\algorithm
\renewcommand\algorithm[1][]{\Algorithm[#1]\setstretch{0.96}}

\usepackage{color,hyperref}
\definecolor{darkblue}{rgb}{0.0,0.0,0.55}
\hypersetup{colorlinks,breaklinks,
            linkcolor=darkblue,urlcolor=darkblue,
            anchorcolor=darkblue,citecolor=darkblue}
            
\definecolor{darkpastelgreen}{rgb}{0.01, 0.75, 0.24}

\renewenvironment{itemize}{
\begin{list}{}{
\setlength{\leftmargin}{4em}
}}{
  \end{list} }
  
\newcommand\blfootnote[1]{%
  \begingroup
  \renewcommand\thefootnote{}\footnote{#1}%
  \addtocounter{footnote}{-1}%
  \endgroup
}

\begin{document}

\title{
Conditional Copula Models for Right-Censored Clustered Event Time Data
}

\author{
Candida Geerdens$^{1}$,
Elif F. Acar$^{2,\ast}$ and
Paul Janssen$^{1}$\\
\small{$^{1}$ Center for Statistics, Universiteit Hasselt}\\
\small{$^{2}$Department of Statistics, University of Manitoba}\\
\footnotesize{$^{\ast}$ Corresponding Author: \href{mailto:elif.acar@umanitoba.ca}{elif.acar@umanitoba.ca}}
}

\date{}

\maketitle

\begin{abstract}

This paper proposes a modelling strategy to infer the impact of a covariate on the dependence structure of right-censored clustered event time data.
The joint survival function of the event times is modelled using a parametric conditional copula whose parameter depends on a cluster-level covariate in a functional way.
We use a local likelihood approach to estimate the form of the copula parameter and outline a generalized likelihood ratio-type test strategy to formally test its constancy. A bootstrap procedure is employed to obtain an approximate $p$-value for the test.
The performance of the proposed estimation and testing methods are evaluated in simulations under different rates of right-censoring and for various parametric copula families, considering both parametrically and nonparametrically estimated margins.
We apply the methods to data from the Diabetic Retinopathy Study to assess the impact of disease onset age on the loss of visual acuity.

\end{abstract}


\blfootnote{Submitted on April 10, 2016.}

\newpage

\section{Introduction}
\label{sec:1}

Many biomedical studies involve clustered time-to-event data, which can be right-censored and which may exhibit strong dependence.
For instance, lifetimes of twins or married couples are often dependent due to shared genetic or environmental factors, and characterizing these dependencies helps making informed decisions in health research.
Other examples include time to failure of matched organs, such as eyes or kidneys, and occurrence times of linked diseases.
In such studies, the data analysis should be directed towards unraveling the within-cluster dependence, or one should at least account for its presence in the applied modelling strategy.
Copula models are well-suited for this task.

Copulas are dependence functions that link together the marginal survival functions to form the joint survival function.
Their use in survival analysis has a long history dating back to \cite{Clayton:1978}, followed by  \cite{Oakes:1982}, \cite{Hougaard:1986}, and more recently, \cite{Shih/Louis:1995} and \cite{Chen/Fan/Douzo/Ying:2010}, among others.

In these papers, the focus is mainly on the unconditional dependence structure of event times and not on the presence of covariates that could provide additional information on the joint survival function.
One exception is \cite{Clayton:1978}, which devotes a section on strategies to include covariates in the association analysis of bivariate failure times and suggests adjusting both the marginal survival functions and the dependence parameter for covariates, but without any elaborate treatment.

\begin{figure}[h!]
\center{
\includegraphics[width=0.99\textwidth]{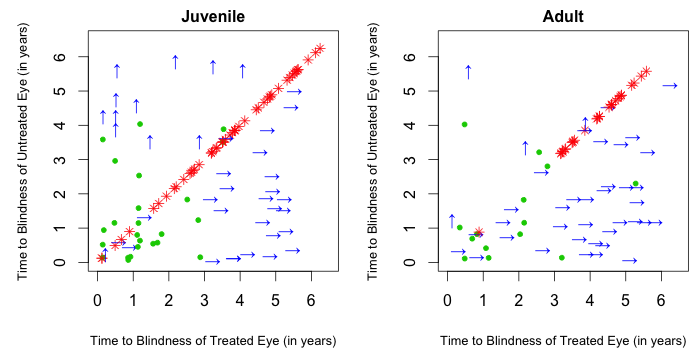}
\caption{Scatter plots of time to blindness (in years) of treated and untreated eyes in the juvenile and adult diabetes groups.
The data points shown via `$\textcolor{darkpastelgreen}{\bullet}$' have events in both eyes;~
`$\textcolor{red}{\ast}$' have censoring in both eyes;~
`$\textcolor{blue}{\rightarrow}$' have only the treated eye censored;~
and `$\textcolor{blue}{\uparrow}$'~ have only the untreated eye censored.
\label{fig:drs1}}
}
\end{figure}

Despite Clayton's suggestion, most commonly used approaches in survival analysis incorporate covariates only in the marginal models, and neglect their potential impact on the association structure.
For instance, in an effort to perform covariate adjustment, \cite{Huster/Brookmeyer/Self:1989} proposed a parametric analysis of paired right-censored event time data in the presence of binary covariates, with an application to diabetic retinopathy data.
In this analysis, the type of diabetes, classified into juvenile or adult groups based on age at onset, is considered as the covariate; and its impact is accounted for only in the marginal models for the time to loss of visual acuity in the laser-treated and untreated eyes, but not in the association structure.
This amounts to an implicit assumption that the dependence parameter is the same for the juvenile and adult groups, which may not be realistic or at least needs to verified.
Note that, based on a visual representation of the data, it is difficult to track whether the dependence parameters
of the two groups differ or not, mainly due to the high rate of right-censoring (see Figure~\ref{fig:drs1}).

While there exists many tools to account for covariates in the marginal survival functions of clustered right-censored time-to-event data, there is a need to extend copula-based models to include covariate information in the association structure.
This paper proposes covariate-adjusted dependence analysis for clustered right-censored event time data using a parametric conditional copula whose parameter is allowed to depend on a cluster-level covariate.
When the latter is binary or discrete with few categories, one can form two or more strata according to the covariate value and fit a copula to each stratum separately.

The impact of a continuous covariate on the dependence parameter is notoriously more difficult to formulate, as it should be specified in functional terms and is typically data specific.
This invites the use of nonparametric techniques for function estimation.

In the case of complete data, i.e. when there is no censoring, nonparametric estimation of the copula parameter function has been previously studied in \cite{Acar/Craiu/Yao:2011} and \cite{Abegaz/Gijbels/Veraverbeke:2012} for parametrically and nonparametrically estimated margins, respectively.
These proposals are built on local likelihood methods \citep{Tibshirani/Hastie:1987} combined with local polynomial estimation \citep{Fan/Gijbels:1996}.
They are, however, not directly applicable to right-censored event times.
The presence of right-censoring in the event times greatly challenges the statistical analysis, and its incorporation in the copula parameter estimation is necessary.
A recent work in this domain is \cite{Ding/Hsieh/Wang:2015}, which proposes a nonparametric estimator for the concordance probability as a function of covariates. However, this approach does not readily generalize to a likelihood-based model formulation.

Here, the first contribution is an extension of the conditional copula framework in \cite{Acar/Craiu/Yao:2011} and \cite{Abegaz/Gijbels/Veraverbeke:2012} to handle right-censored event time data, for both parametrically and nonparametrically estimated marginal survival functions. For the former, we focus on the Weibull model as employed in \cite{Huster/Brookmeyer/Self:1989}, and for the latter we consider the Beran's estimator \citep{Beran:1981}.

The second contribution is a test strategy for the constancy of the conditional copula parameter across the range of a cluster-level covariate. In the case of a discrete covariate, one can employ the traditional likelihood ratio test to assess whether the dependence parameters for different strata can be deemed the same.
However, for a continuous covariate, one is required to test the constancy of the whole dependence parameter function.
Here, this is achieved by adopting the test strategy in \cite{Acar/Craiu/Yao:2013}. The test is built on the generalized likelihood ratio statistic of \cite{Fan/Zhang/Zhang:2001} for testing a parametric or a nonparametric null hypothesis versus a nonparametric alternative hypothesis.
For conditional copulas with complete data, \cite{Acar/Craiu/Yao:2013} derived the asymptotic null distribution of the test statistic and used it to obtain a decision rule.
The presence of right-censoring complicates the development of the asymptotic null distribution. Therefore, we alternatively propose a bootstrap procedure to obtain an approximate p-value for the test.

The proposed estimation and testing methods are detailed in Sections~\ref{sec:2} and \ref{sec:3}, respectively.
Section~\ref{sec:4} contains the results from our simulations under different rates of right-censoring and for various parametric copula families, considering both parametrically and nonparametrically estimated margins.
In Section~\ref{sec:5}, we revisit the diabetic retinopathy data and assess the impact of age at onset on the time to loss of visual acuity.
The paper concludes with a brief discussion. 
The bootstrap algorithms are provided in the Appendix. 
Part of the simulation and data analysis results are collected in the Supplemental Material.

\vspace*{-8pt}
\section{Conditional Copula Model for Right-Censored Event Time Data} 
\label{sec:2}

In this section, we introduce the notation and describe the proposed conditional copula approach for right-censored clustered event times.
To ease the presentation, we focus on the bivariate setting. However, the results can be extended to settings with clusters of higher (but equal) size using a multivariate copula.

Let $(T_1, T_2)$ be a vector of bivariate event times, and let $X$ be a continuous cluster-level covariate.
Then, for each $x$ in the support of $X$, the conditional joint survival function $S_X(t_1,t_2|x)=P(T_1>t_1,T_2>t_2|X=x)$ of the vector $(T_1,T_2)|X=x$ has a unique representation (Patton, 2002) given by
\beq
\label{eq1}
S_X(t_1,t_2 \mid X= x) = \mathbb{C}_{X}\left( S_{1\mid x}(t_1 \mid x), S_{2 \mid x}(t_2 \mid x) \mid X=x \right),
\eeq
where $\mathbb{C}_{X}$ is the conditional copula of the event times, and $S_{k \mid x}(t_k|x)=P(T_k>t_k|X=x)$ is the conditional marginal survival function of $T_k|X=x$ ($k=1,2$).

A major complication in fitting the model in \eqref{eq1} is that for right-censored data, the true event time is not always recorded, but instead, a lower time, called the censoring time, is observed.
Let $(C_1, C_2)$ be a vector of bivariate censoring times, independent of $(T_1,T_2)$.
We observe the minima $(Y_1,Y_2)= \left(\min\{T_1,C_1\},\min\{T_2,C_2\}\right)$, together with the corresponding censoring indicators $(\delta_1,~\delta_2)=(I\{T_1\leq C_1\},~I\{T_2\leq C_2\})$. In the special case where the same censoring time applies to all members of a cluster, we have  $C= C_1=C_2$, a situation referred to as univariate censoring.

Given a random sample $\{(Y_{1i}, Y_{2i}, \delta_{1i}, \delta_{2i}, X_i)\}_{i=1}^n$, the fitting of the model in \eqref{eq1} is typically performed in two-stages; first for the conditional marginal survival functions $S_{k\mid x}$, and second for the conditional copula $\mathbb{C}_{X}$.
To estimate $S_{k\mid x}$, one can employ parametric or nonparametric strategies.
These are outlined briefly in Section \ref{sec:2.1}.
We then describe the proposed nonparametric strategy for fitting the conditional copula in Section \ref{sec:2.2}. The details on the bandwidth selection for the nonparametric methods are given in Section \ref{sec:2.3}.

\vspace*{-8pt}
\subsection{Estimation of the conditional marginal survival functions} 
\label{sec:2.1}

In the case of parametric conditional margins, such as Weibull, we have $ S_{k\mid x}(t_k \mid x) =  S_{k\mid x}(t_k \mid x, \bs{{\gamma}}_k )$, with  $\bs{{\gamma}}_k$ an unknown parameter vector ($k=1,2$).
One can then use maximum likelihood estimation to obtain
$$\widehat{S}_{k\mid x}(t_k \mid x) =  S_{k\mid x}(t_k \mid x, \bs{\hat{\gamma}}_k ),$$
where
$
\bs{\hat{\gamma}}_k  = \argmax_{\bs{\gamma}_k}{  \sum_{i=1}^n  \delta_{ki} \ln f_{k\mid x}(Y_{ki} | X_i , \bs{\gamma}_k) + (1- \delta_{ki}) \ln S_{k\mid x}(Y_{ki} | X_i , \bs{\gamma}_k)},
$
is the maximum likelihood estimate of the vector of marginal parameters ($k=1,2$).

In the absence of a suitable parametric model, the conditional margins can be estimated nonparametrically using the Beran's estimator (Beran, 1981), also called the conditional Kaplan-Meier estimator, which takes the form
$$
\widetilde{S}_{k\mid x}(t_k \mid x)   = \prod_{Y_{ki} \leq t_k, \delta_{ki} = 1}   \left(       1-     \dfrac{ w_{nki}  (x; h_{k})    } {  \sum_{j=1}^{n}  I\{Y_{kj}\geq Y_{ki}\} w_{nkj} (x; h_{k}) }   \right)
$$
The weights $w_{nki}$ are typically defined as 
$$
 w_{nki}  (x; h_{k}) = \dfrac{  K_{h_k}\left( X_{i} - x \right) }{\sum_{j=1}^n  K_{h_k}\left( X_{j}- x \right) },
$$
where $K_{h_k} (\cdot)= K(\cdot /h_k) /h_k $, with $K$ the kernel function and $h_k$ the bandwidth parameter ($k=1,2$).

\vspace*{-8pt}
\subsection{Estimation of the conditional copula} 
\label{sec:2.2}

Given the estimated margins, and assuming that for each value of $x$, the conditional copula $\mathbb{C}_{X}$ belongs to the same parametric copula family, one can fit $\mathbb{C}_{X}$ within the likelihood framework.
In this case, the impact of a covariate is considered to be solely on the strength of dependence, which is captured by the copula parameter $\theta(X)$ of $\mathbb{C}_{X}$.

Due to the restricted parameter range of some copula families, instead of directly modelling $\theta(X)$, we consider the reparametrization $ \theta(x) = g^{-1}(\eta(x))$, where $\eta(\cdot)$ is called the \emph{calibration function} and $g^{-1}:  \mathbb{R} \rightarrow \Theta $ is a prespecified inverse-link function with $\Theta$ being the parameter space of a given copula family. For some commonly used copula families, the inverse link functions are provided in Table \ref{table:Copulas}.

\begin{table}[h]  \footnotesize
\begin{centering}
\caption{Inverse link functions and Kendall's tau conversions for some copula families.}
\label{table:Copulas}
\begin{tabular}{ l l  c  c c  c c  c}
&Family    &&        $\mathbb{C}(u_{1},u_{2})$       &&    $\theta $ & $g^{-1}(\eta)$ & $\tau$
\\  [1ex] \hline
&Clayton  &&     $ (u_{1}^{-\theta}+u_{2}^{-\theta}-1)^{-\frac{1}{\theta}}  $  &&   $(0,\infty)$    & $\exp(\eta)$  &$\frac{\theta}{\theta +2}$
\\ [2ex]
&Frank  && $- \frac{1}{\theta}  \ln \Big\{  1+ \frac{(e^{-\theta u_{1}}-1)    (e^{-\theta u_{2}}-1)}{   e^{-\theta }-1} \Big\}   $  &&   $(-\infty,\infty) \setminus \{0\}$ &  $\eta$  &   $1+ \frac{4}{\theta} [D_{1}(\theta)-1]$
\\ [2ex]
&Gumbel  && $  \exp \Big\{ - [(-\ln u_{1})^{\theta} + (- \ln u_{2})^{\theta} ]^{\frac{1}{\theta}} \Big\}  $ && $[1,\infty) $ & $\exp(\eta)+1$ &  $1- \frac{1}{\theta}$
\\ [0.5ex]
\hline
\end{tabular}
\end{centering}
\\
{\footnotesize where $D_{1}(\theta) = \frac{1}{\theta}  \int_{0}^{\theta} \frac{t}{e^{t}-1} dt $ is the Debye function.}
\end{table}

Letting $U_k \equiv S_{k\mid x}(\cdot \mid x)$ for $k=1,2$, the model in \eqref{eq1} becomes
\beq
\label{eq2}
\left(U_{1},U_{2}\right)  \mid X= x  \sim \mathbb{C}_{X} \left( u_{1},u_{2} \mid  g^{-1}(\eta(x)) \right).
\eeq
Hence, the loglikelihood function takes the form \citep{Shih/Louis:1995, Massonnet:2009}
\beq
\label{eq3}
 \displaystyle \sum_{i=1}^n \ell \left(  g^{-1}(\eta(X_i)), {U}_{1i}, {U}_{2i}     \right),
\eeq
where
\begin{eqnarray*}
\ell (v, u_1,u_2 ) &=& (1-\delta_{1}) (1-\delta_{2}) \ln \mathbb{C}_{X}(u_1,u_2 \mid v ) \\
&& + \quad \delta_{1} (1-\delta_{2}) \ln \left[\frac{\partial \mathbb{C}_{X}(u_1,u_2 \mid v )}{\partial u_1} \right] \\
&& + \quad  (1-\delta_{1}) \delta_{2} \ln \left[\frac{\partial \mathbb{C}_{X}(u_1,u_2 \mid v)}{\partial u_2} \right] \\
&& + \quad  \delta_{1} \delta_{2} \ln \left[\frac{\partial^2 \mathbb{C}_{X}(u_1,u_2 \mid v)}{\partial u_1 \partial u_2} \right].
\end{eqnarray*}
Note that due to right-censoring, the loglikelihood contributions of the data vectors are non-trivial,  i.e., they involve the copula function $\mathbb{C}_{X}$ as well as its first and second order derivatives.

To fit the conditional copula, one can use maximum likelihood estimation by specifying a parametric form for $\eta(X)$. However, as noted before, the impact of the covariate on the dependence strength is difficult to predetermine in most applications. Therefore, it is often advised to employ nonparametric strategies \citep{Acar/Craiu/Yao:2011, Abegaz/Gijbels/Veraverbeke:2012}.

Suppose that  $\eta(\cdot)$ is sufficiently smooth, i.e., $\eta(\cdot)$ has the ($p+1$)th derivative at the point $x$. Then, for a covariate value $X_i$ in a neighborhood of $x$, we can approximate $\eta(X_i)$ by a Taylor expansion of order $p$:
\begin{eqnarray*}
\eta(X_i) &\approx& \eta(x) + \eta^{(1)}(x) (X_i - x) + \ldots + \displaystyle \frac{\eta^{(p)}(x)}{p!} (X_i - x)^p \\
& \equiv & \beta_{0,x} + \beta_{1,x} (X_i - x) + \ldots + \beta_{p,x} (X_i - x)^p \equiv {\bs{x}}_{i,x}^{T}~\bs{\beta}_x
\end{eqnarray*}
where ${\bs x}_{i, x}= (1, X_{i}-x,\ldots,  (X_{i}-x)^{p}  ) ^{T} $ and
$\bs{\beta}_x = (\beta_{0,x},\beta_{1,x},\ldots,\beta_{p,x})^{T}$ with $\beta_{r,x}=\eta^{(r)}(x) / r!$.
We then estimate $\bs{\beta}_x$ by maximizing a local version of \eqref{eq3}, which is given by
\beq
\label{eq4}
\sum_{i=1}^n  \ell \left( g^{-1}( \bs{x}_{i,x}^{T}  \bs{\beta}_x),  {U}_{1i},  {U}_{2i}  \right) \;  K_{h_\mathbb{C}}(X_i - x),
\eeq
where $K_{h_\mathbb{C}} (\cdot)= K(\cdot /h_\mathbb{C}) /h_\mathbb{C} $, with $K$ the kernel function, $h_\mathbb{C}$ the bandwidth parameter and ${U}_{ki} \equiv {S}_{k\mid x}(Y_{ki} \mid X_i) $ ($k=1,2$).

In practice, the conditional survival margins  $U_{ki}$ in \eqref{eq4} are replaced by either parametric estimates,  $\widehat{U}_{ki} \equiv \widehat{S}_{k\mid x}(Y_{ki} \mid X_i) $, or nonparametric estimates $\widetilde{U}_{ki} \equiv  \widetilde{S}_{k\mid x}(Y_{ki}~\mid~X_i)$ ($k=1,2$).
The resulting local maximum likelihood estimates are denoted by $\bs{\widehat{\beta}}_x =  (\hat{\beta}_{0,x},\hat{\beta}_{1,x}, \ldots, \hat{\beta}_{p,x})^{T} $ and $\bs{\widetilde{\beta}}_x= (\tilde{\beta}_{0,x},\tilde{\beta}_{1,x}, \ldots, \tilde{\beta}_{p,x})^{T}$, respectively.
From these, one can obtain $\hat{\eta}(x) = \hat{\beta}_{0,x}$ and $\tilde{\eta}(x) = \tilde{\beta}_{0,x}$, which in turn yield the estimates of the copula parameter at covariate value $x$  via $\hat{\theta}(x) = g^{-1} ( \hat{\eta}(x) )$  and $\tilde{\theta}(x) =  g^{-1} ( \tilde{\eta}(x) )$.

\vspace*{-8pt}
\subsection{Bandwidth selection} 
\label{sec:2.3}

The choice of the bandwidth parameter $h_\mathbb{C} $ plays an important role in the local likelihood estimation given in \eqref{eq4}.
If the conditional marginal survival functions are estimated parametrically, the leave-one-out cross-validated loglikelihood criterion in \cite{Acar/Craiu/Yao:2011} can be employed to obtain a data-driven bandwidth. In this case, we select the bandwidth value that maximizes the leave-one-out cross-validated loglikelihood function
\beq
\label{eq:5}
B(h_\mathbb{C} ) = \displaystyle \sum_{i=1}^n   \ell \left( \widehat{\theta}^{(-i)}_{h_\mathbb{C}}(X_i) ,   \widehat{U}_{1i},  \widehat{U}_{2i}  \right),
\eeq
where $\widehat{\theta}^{(-i)}_{h_\mathbb{C}}(X_i)$ is the estimated copula parameter at the bandwidth value ${h_\mathbb{C}}$, obtained by leaving the $i^{\rm{th}}$ data point ($\widehat{U}_{1i},\widehat{U}_{2i}, \delta_{1i}, \delta_{2i}, X_i$) out and using the remaining data points ($\widehat{U}_{1j},\widehat{U}_{2j}, \delta_{1j}, \delta_{2j}, X_j$) with $j = 1,\ldots, i-1, i+1, \ldots, n$.

If the Beran's estimator is used to obtain the  conditional marginal survival functions, the bandwidth selection involves two additional parameters, $h_1$ and $h_2$.
Since the model fitting is performed in two stages, one could first choose $h_1$ and $h_2$, separately, and then determine $h_\mathbb{C}$.
However, the few available bandwidth selectors for the Beran's estimator are either not easy to implement or not data-driven \citep{VanKeilegom:1998, Gang/Datta:2001, Demin/Chimitova:2014}.
Therefore, we propose to choose the bandwidth values $(h_1, h_2, h_\mathbb{C})$ jointly,
exploiting the loglikelihood framework. This amounts to maximizing the following function
\beq
\label{eq:6}
B^{\ast}(h_1, h_2, h_\mathbb{C} ) = \displaystyle \sum_{i=1}^n   \ell \left( \widetilde{\theta}^{(-i)}_{h_\mathbb{C}}(X_i) ,   \widetilde{U}_{1i (h_1)},  \widetilde{U}_{2i (h_2)}  \right).
\eeq
Here $\widetilde{U}_{ki (h_k)}$ denotes the nonparametric estimate of the $k^{\rm{th}}$ conditional margin at the bandwidth value $h_k$ ($k=1,2$), and $\widetilde{\theta}^{(-i)}_{h_\mathbb{C}}(X_i)$ is the leave-one-out cross-validated copula parameter estimate for the $i^{\rm{th}}$ observation based on the fitted conditional margins and using the bandwidth value ${h_\mathbb{C}}$.
Note that, due to joint selection of $(h_1, h_2, h_\mathbb{C})$,  the estimates of the conditional margins and the copula parameter depend on all three bandwidth values. Nevertheless, for notational simplicity, we use only the corresponding bandwidth value in each component.

\vspace*{-8pt}
\section{Generalized Likelihood Ratio Test} 
\label{sec:3}

A relevant question in applications is whether a covariate has a significant impact on the underlying dependence structure of the clustered event times. This is equivalent to testing the constancy of the conditional copula parameter as formalized by
\beqn
\label{eq:null}
\rm{H}_0 : \theta(\cdot) \in \mathfrak{f}_c   \qquad \quad  \text{versus} \qquad \quad \rm{H}_1 : \theta(\cdot) \notin \mathfrak{f}_c,
\eeqn
where $\mathfrak{f}_c = \{ \theta(\cdot): \exists \; \theta_0 \in \Theta \mbox{ such that } \theta(X)=\theta_0,  \; \; \forall X\in \mathcal{X}  \}$ is the set of all constant functions on $X\in \mathcal{X}$.
Note that the null hypothesis corresponds to the so-called simplifying assumption in pair-copula constructions \citep{HobaekHaff/Aas/Frigessi:2010, Acar/Genest/Neslehova:2012}.

The hypothesis testing problem in \eqref{eq:null} can be evaluated through the difference \citep{Acar/Craiu/Yao:2013}
$$
\lambda_n = \sup_{\theta(\cdot) \notin \mathfrak{f}_c } \{\mathbb{L}_n (\rm{H}_1) \}  - \sup_{\theta(\cdot) \in \mathfrak{f}_c}  \{\mathbb{L}_n (\rm{H}_0)\},
$$
where
$$
\mathbb{L}_n ({\rm{H}}_0) =\sum_{i=1}^n  \ell \left( \theta_0 , U_{1i}, U_{2i} \right)
 \qquad \text{and} \qquad
\mathbb{L}_n ({\rm{H}}_1) = \sum_{i=1}^n  \ell \left( {\theta}(X_i), U_{1i}, U_{2i} \right).
$$
The statistic $\lambda_n$ is referred to as the \emph{generalized likelihood ratio} (GLR), and differs from the canonical likelihood ratio in that the model under the alternative hypothesis is specified nonparametrically. Hence, the distribution of the test statistic depends on the bandwidth parameter and the kernel function used in the nonparametric estimation.
Further, the presence of right-censoring complicates the loglikelihood expressions, making the assessment of the asymptotic null distribution of the test statistic quite cumbersome.
Even when available, the convergence of the null distribution to the asymptotic distribution might be slow; hence a bootstrap estimate is typically used in finite samples to approximate the null distribution \citep{Fan/Zhang:2004, Fan/Jiang:2005}.
Here, we follow a similar strategy and propose a bootstrap algorithm to obtain an approximate p-value for the test.
We distinguish two cases according to whether the conditional marginal survival functions are estimated parametrically or nonparametrically.

When the conditional marginal survival functions are estimated parametrically, the supremum of the loglikelihood function under the null hypothesis is given by
$$
\mathbb{L}_n ({\rm{H}}_0, \widehat{\theta}_0)= \sum_{i=1}^n \ell \left( \widehat{\theta}_0,  \widehat{U}_{1i}, \widehat{U}_{2i} \right),
$$
where $ \widehat{\theta}_0$ is the maximum likelihood estimator of the constant conditional copula parameter $\theta_0$ based on observations $(\widehat{U}_{1i}, \widehat{U}_{2i}, \delta_{1i}, \delta_{2i})$, $i=1,\ldots,n$.
For the alternative hypothesis, we use the local likelihood estimator $\widehat \theta_{h_\mathbb{C}}$ at each covariate value (Section \ref{sec:2.2}) and obtain
$$
\mathbb{L}_n ({\rm{H}}_1,\widehat \theta_{h_\mathbb{C}} ) =  \sum_{i=1}^n  \ell \left( \widehat{\theta}_{h_\mathbb{C}} (X_i),  \widehat{U}_{1i}, \widehat{U}_{2i}  \right),
$$
where $\widehat{\theta}_{h_\mathbb{C}}$ is the estimated copula parameter obtained by maximization of \eqref{eq:5} with bandwidth $h_\mathbb{C}$, the optimal bandwidth value. The generalized likelihood ratio statistic then becomes
$$
\lambda_n (h_\mathbb{C}) =    \sum_{i=1}^n  \ell \left( \widehat{\theta}_{h_\mathbb{C}} (X_i),  \widehat{U}_{1i}, \widehat{U}_{2i}  \right) -  \sum_{i=1}^n \ell \left(  \widehat{\theta}_0,  \widehat{U}_{1i}, \widehat{U}_{2i} \right).
$$
To obtain the null distribution of $\lambda_n (h_\mathbb{C})$, we use the bootstrap procedure outlined in Algorithm \ref{alg:1} (see the Appendix) 
\citep{DavisonHinkley:1997, Geerdens2013, BGeerdens2015}.

In the case of nonparametrically estimated conditional margins, a similar strategy is followed to obtain
$$
\lambda_n (h_1, h_2, h_\mathbb{C}) =    \sum_{i=1}^n  \ell \left( \widetilde{\theta}_{h_\mathbb{C}} (X_i),  \widetilde{U}_{1i}, \widetilde{U}_{2i}  \right) -  \sum_{i=1}^n \ell \left(  \widetilde{\theta}_0,  \widetilde{U}_{1i}, \widetilde{U}_{2i} \right),
$$
where $\widetilde{\theta}_0$ is the maximum likelihood estimator of the constant conditional copula parameter $\theta_0$ based on observations $(\widetilde{U}_{1i}, \widetilde{U}_{2i}, \delta_{1i}, \delta_{2i})$, $i=1,\ldots,n$. The latter are obtained using
$h_1$ and $h_2$, the optimal bandwidth values maximizing \eqref{eq:6}, jointly with the bandwidth value $h_\mathbb{C}$.
For the alternative model, we use the local likelihood estimator $\widetilde{\theta}_{h_\mathbb{C}}$ at each covariate value (Section \ref{sec:2.2}).
To obtain the null distribution of $\lambda_n (h_1, h_2, h_\mathbb{C})$, we employ Algorithm  \ref{alg:2} (see the Appendix),
which differs from Algorithm \ref{alg:1} mainly in that $(\widehat{U}_{1i}, \widehat{U}_{2i})$ and $\widehat{S}_{k|x}$ are replaced by $(\widetilde{U}_{1i}, \widetilde{U}_{2i})$ and $\widetilde{S}_{k|x}$.
In the bootstrap, the bandwidth values are taken to be the same as the ones obtained for the original data.

In both algorithms, the null distribution of the test statistic is only approximate. Each bootstrap sample, although generated using the constant conditional copula parameter value under the null hypothesis, involves additional variation/uncertainty arising from the estimation of the conditional marginal survival functions.
This variation is more pronounced when the Beran's estimator is used (see Section \ref{sec:4}).

\vspace*{-8pt}
\section{Simulation Results} 
\label{sec:4}
We investigate the finite sample performance of the proposed estimation and testing methods in a simulation study.
We consider the Clayton, Frank and Gumbel copulas, with dependence structures given by the following scenarios:
\ben
\item[] Constant model:~~~~$ \tau(X) = 0.6$
\item[] Convex model:~~~~~~$\tau(X) = 0.1(X-3)^2 +0.3$
\item[] Concave model:~~~~~$\tau(X) = -0.1 (X-3)^2 +0.7$
\een

The models for the covariate effect are specified in terms of Kendall's tau to allow comparisons across different copulas. The conversions between the copula parameter $\theta$ and Kendall's tau $\tau$ {}{are} given in Table \ref{table:Copulas} for the considered copulas.
In the constant model, the covariate has no effect on the strength of dependence, while in the convex and concave models, the covariate effect has the respective form with Kendall's tau varying from $0.3$ to $0.7$.

Under each scenario, we generate the copula data $\{ (U_{1i},U_{2i} \mid  X_{i}): i=1,2,\ldots,n  \}$ as outlined in \cite{Acar/Craiu/Yao:2011}. That is, we first obtain covariate values $X_{i}$ from Uniform~$(2, 5)$. Then, for each $i=1,2,\ldots,n$, we calculate the corresponding Kendall's tau $\tau_i \equiv\tau(X_i)$ and the corresponding copula parameter $\theta_i \equiv \theta(X_i)$.
To obtain the event times, we apply the inverse-cdf method to the copula data using the Weibull model with parameters $\lambda_T=0.5$, $\rho_T=1.5$, and $\beta_T=0.8$. To introduce right-censoring, we generate the (univariate) censoring times from the Weibull model with parameters $\lambda_C=1.5$ and $\rho_C=1.5$ for the case of low censoring (approximately 20\% censoring rate), and with parameters $\lambda_C=1.5$ and $\rho_C=0.5$ for the case of moderate censoring (approximately 50\% {censoring rate). The observed data are then given by the minima of the event times and the censoring times.
We also consider non-censored event time data to assess the impact of censoring on the results.

We estimate the conditional marginal survival function parametrically, using the Weibull model, and nonparametrically, using the Beran's estimator. Based on the resulting estimates, we perform local linear estimation under the correct copula and obtain the estimates of the calibration function. The local weights are defined using the Epanechnikov kernel.
For the bandwidth parameter(s), we consider $6$ candidate values, ranging from $0.3$ to $3$, equally spaced on a logarithmic scale. 
The bandwidth selection for the Beran's estimator is based on a comparison with the true marginal survival function, while the bandwidth selection for the conditional copula estimation is based on the cross-validated loglikelihood criterion, and this for each simulated data set.}
All results reported are based on the local calibration estimates at the chosen optimal bandwidth. These calibration estimates are converted into the copula parameter via the link functions in Table~\ref{table:Copulas}.

We evaluate the estimation strategy through the integrated Mean Square Error (IMSE) along with the integrated square Bias (IBIAS$^{2}$) and integrated Variance (IVAR), given by
\beqn
&& \text{IBIAS}^{2}(\hat{\tau}) =   \int_{\mathcal{X}}  \big[   E [  \hat{\tau}(x)]  - \tau(x) \big]^{2} \; dx, \nonumber \\
&& \text{IVAR}(\hat{\tau}) = \int_{\mathcal{X}} E  \big[ \{ \hat{\tau}(x) - E [ \hat \tau(x)]  \}^{2} \big] \; dx, \nonumber \\
&& \text{IMSE}(\hat{\tau}) = \int_{\mathcal{X}}  E \big[ \{ \hat{\tau}(x) - \tau(x) \}^{2} \big] \; dx \; = \; \text{IBIAS}^{2}(\hat{\tau})  + \text{IVAR}(\hat{\tau}). \nonumber
\eeqn

These quantities are approximated by a Riemann sum over a grid of points $\{2,2.1,\ldots,4.9,5\}$ in the covariate range.
We use the Kendall's tau scale to compare the performance of the local linear estimator across different copulas and at different censoring rates.
Under each scenario, we conduct experiments with sample sizes $n=250$ and $500$, each replicated $M=500$ times.
The results under the Clayton copula are displayed in Table~\ref{MISE_Clayton}, while the results under the Frank and Gumbel copulas are deferred to the Supplemental Material.

\begin{table}[h!]
\footnotesize
\caption{Integrated Squared Bias, Integrated Variance and Integrated Mean Square Error (multiplied by 100) of the Kendall's tau estimates under the Clayton copula.}
\label{MISE_Clayton}
\centering
\vspace{0.1in}
\begin{tabular}{cccccccccccccc}
\toprule \\ [-3.5ex]
&&&&& \multicolumn{6}{c}{} &&& \\[-1.75ex]
&&&&  \multicolumn{3}{c}{Parametric Margins}&&&\multicolumn{3}{c}{Nonparametric Margins}  \\[0.5ex] 
\cmidrule(r){5-8} \cmidrule(r){10-13}
 & censoring rate & n && IBIAS$^{2}$ & IVAR & IMSE &&& IBIAS$^{2}$ & IVAR & IMSE \\
\midrule
\multirow{6}{*}{Constant}
& 0 \% & 250  && 0.002 & 0.605 & 0.607 &&& 0.489 & 1.246 & 1.735\\
& & 500       && 0.000 & 0.304 & 0.304 &&& 0.207 & 0.550 & 0.757 \\
& 20 \% & 250 && 0.002 & 0.764 & 0.766 &&& 0.717 & 1.914 & 2.631 \\
& & 500       && 0.001 & 0.392 & 0.393 &&& 0.318 & 0.798 & 1.116 \\
& 50 \% & 250 && 0.005 & 1.261 & 1.266 &&& 1.299 & 3.664 & 4.963 \\
& & 500       && 0.005 & 0.638 & 0.643 &&& 0.675 & 1.547 & 2.222
\\ [0.8ex] \midrule
\multirow{6}{*}{Convex}
& 0 \% & 250  && 0.097 & 1.383 & 1.479 &&& 0.711 & 1.941 & 2.652\\
& & 500       && 0.056 & 0.681 & 0.737 &&& 0.317 & 0.922 & 1.239 \\
& 20 \% & 250 && 0.117 & 1.950 & 2.066 &&& 1.003 & 2.780 & 3.783 \\
& & 500       && 0.060 & 0.964 & 1.024 &&& 0.554 & 1.360 & 1.914 \\
& 50 \%& 250  && 0.175 & 3.303 & 3.478 &&& 1.516 & 4.623 & 6.140 \\
& & 500       && 0.077 & 1.612 & 1.688 &&& 0.913 & 2.044 & 2.957
\\ [0.8ex] \midrule
\multirow{6}{*}{Concave}
& 0 \% & 250  && 0.084 & 0.908 & 0.992 &&& 0.736 & 1.677 & 2.413\\
& & 500       && 0.041 & 0.477 & 0.518 &&& 0.332 & 0.754 & 1.085 \\
& 20 \% & 250 && 0.097 & 1.113 & 1.210 &&& 0.924 & 2.446 & 3.371 \\
& & 500       && 0.056 & 0.571 & 0.627 &&& 0.475 & 1.059 & 1.534 \\
& 50 \%& 250  && 0.149 & 1.805 & 1.954 &&& 1.586 & 4.081 & 5.667 \\
& & 500       && 0.101 & 0.888 & 0.989 &&& 0.890 & 1.953 & 2.843
 \\ [0.8ex]
\bottomrule
\end{tabular}
\end{table}

\begin{figure}[p!]
\begin{center}
{\includegraphics[width=7cm,height=6.5cm]{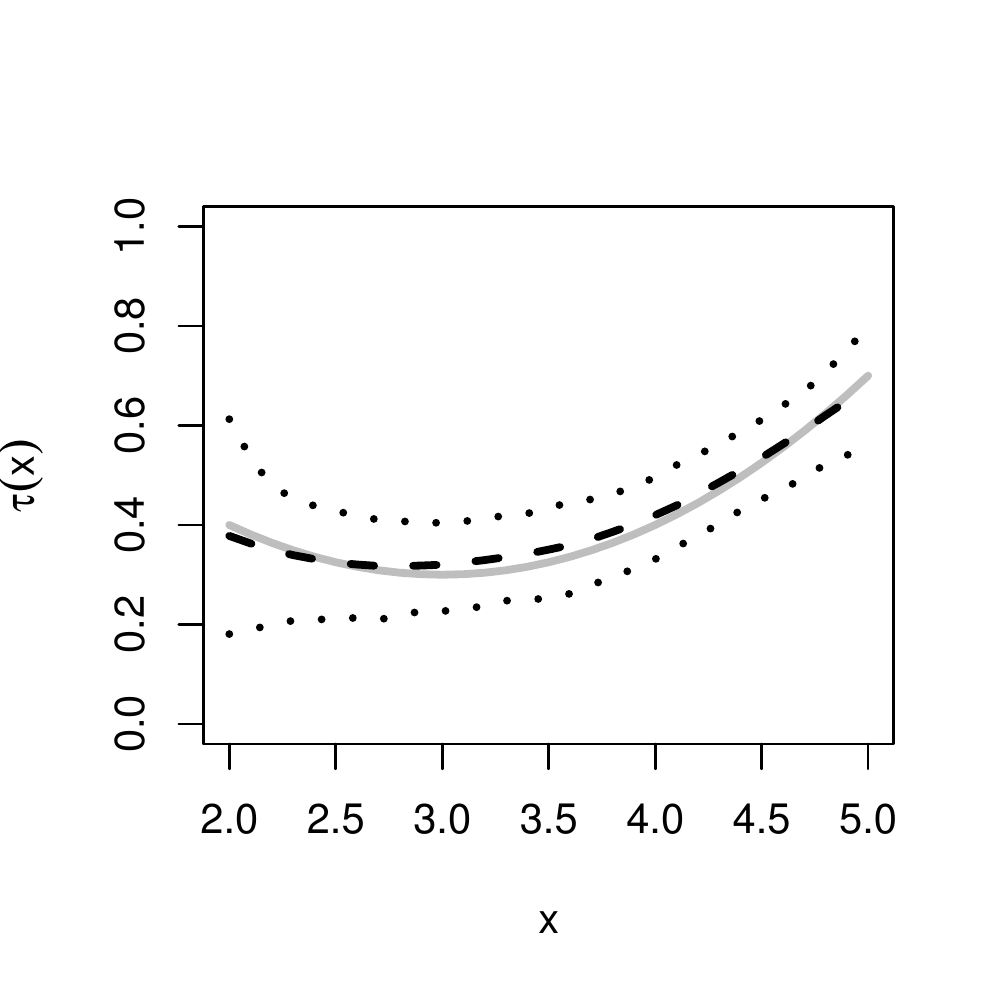}}
{\includegraphics[width=7cm,height=6.5cm]{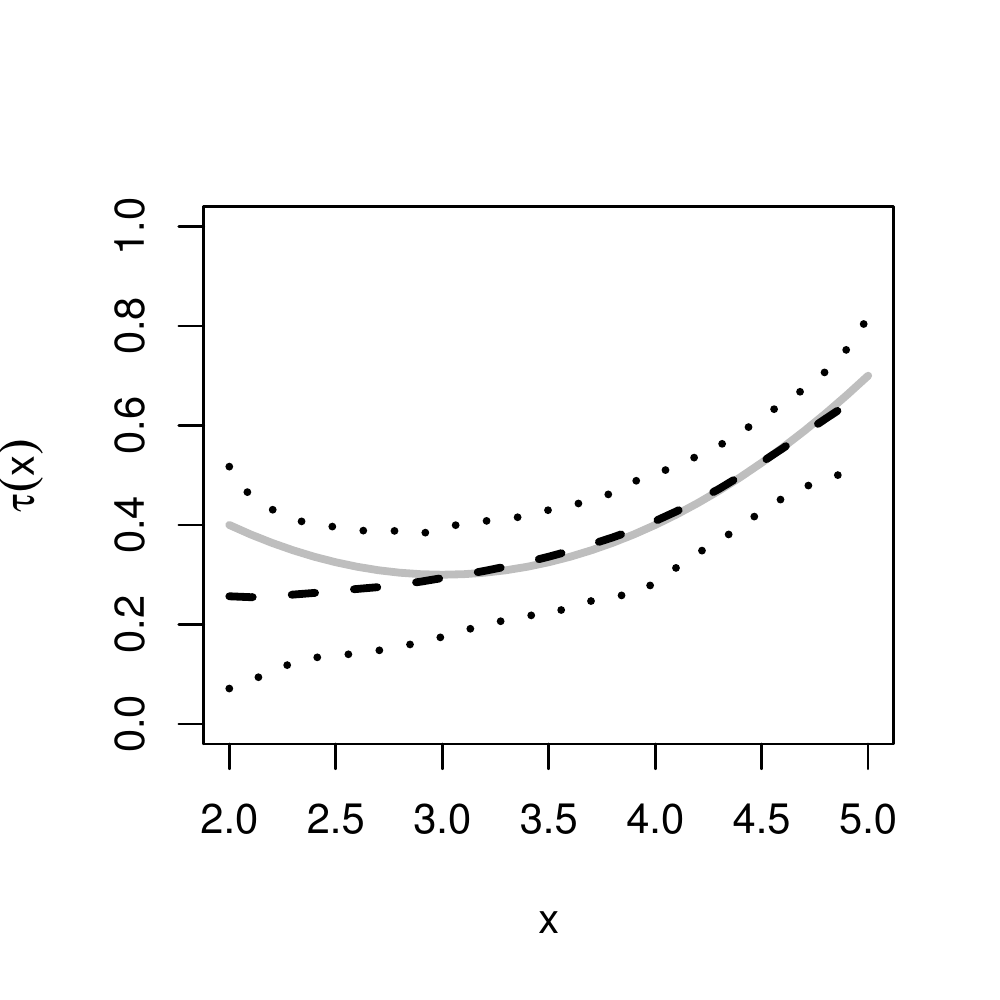}} \\
\vspace{-1cm}
{\includegraphics[width=7cm,height=6.5cm]{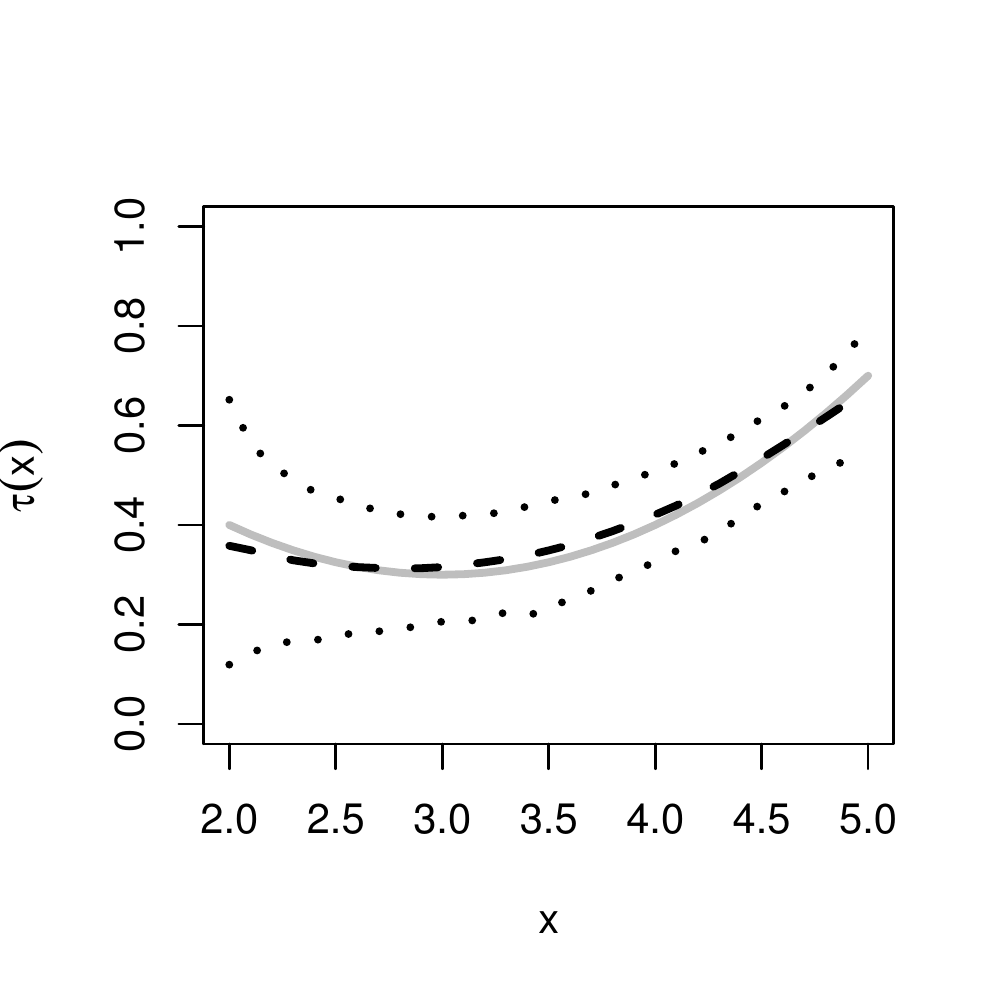}}
{\includegraphics[width=7cm,height=6.5cm]{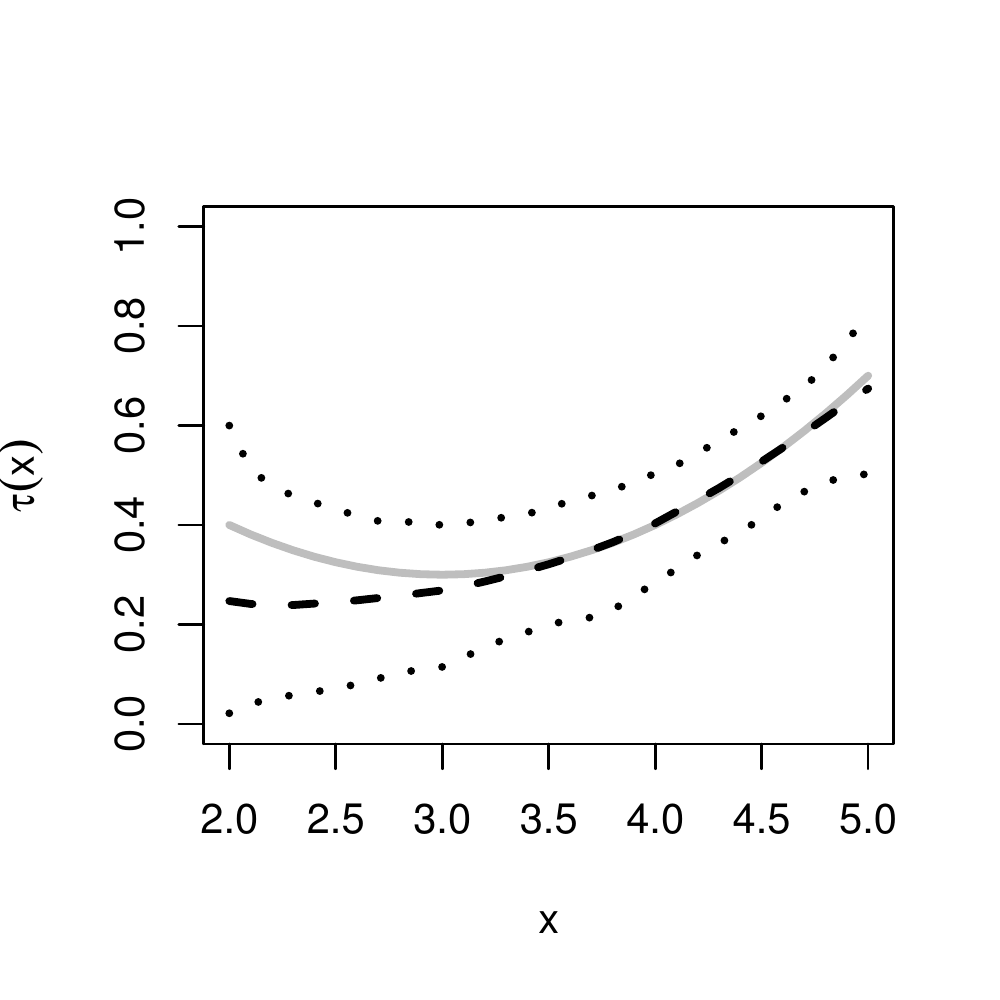}} \\
\vspace{-1cm}
{\includegraphics[width=7cm,height=6.5cm]{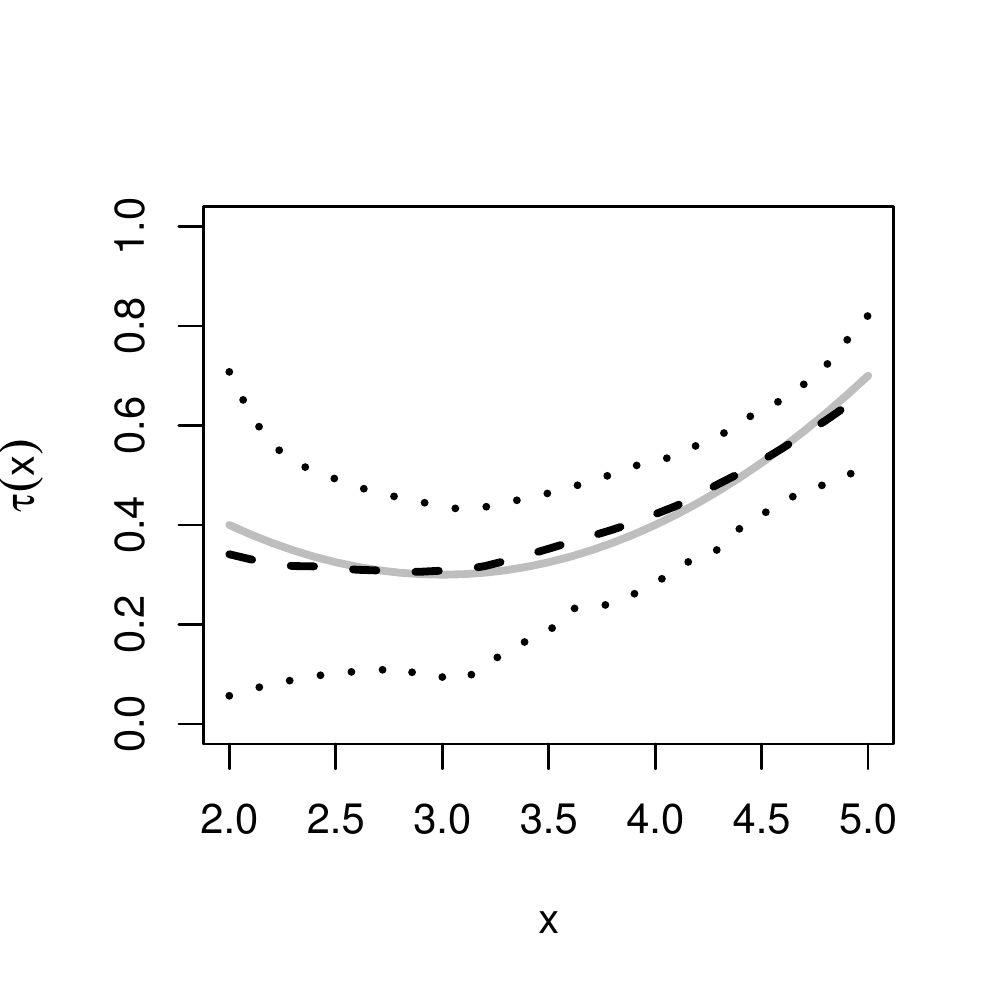}}
{\includegraphics[width=7cm,height=6.5cm]{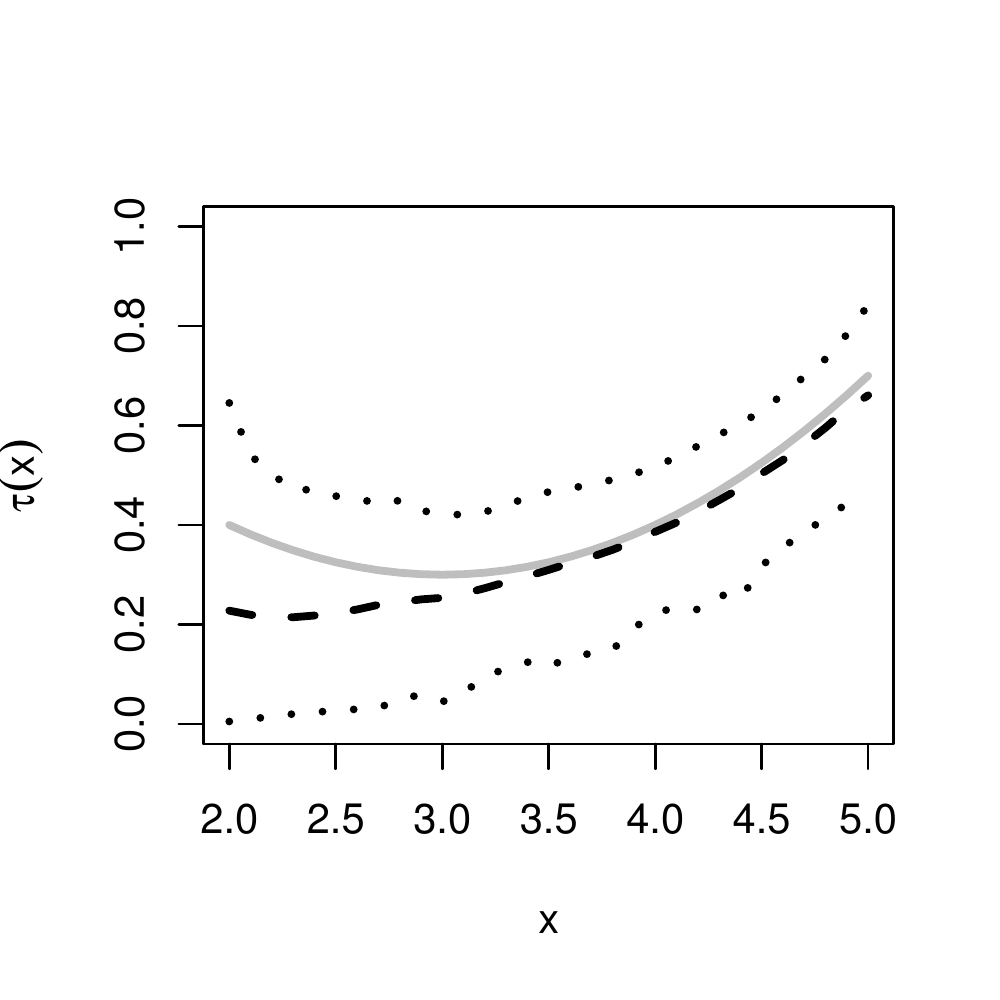}}
\caption{The Kendall's tau estimates for the convex model with sample size $n=250$ under the Clayton copula. Parametric margins (left panel) and nonparametric margins (right panel) - no censoring (top panel), $20\%$ of censoring (middle panel) and $50\%$ of censoring (bottom). 
The grey line represents the true Kendall's tau; the dashed line represents the Kendall's tau estimates averaged over 500 Monte-Carlo samples; and the dotted lines represents the $5^{\rm{th}}$ and $95^{\rm{th}}$ percentiles of the Kendall's tau estimates, respectively.}
\label{fig:estimation_Clayton}
\end{center}
\end{figure}

From Table~\ref{MISE_Clayton}, it can be seen that the estimation performance deteriorates with increasing censoring rate. Since right-censoring causes loss of information, this result is to be expected. 
Further, Table~\ref{MISE_Clayton} shows that the parametrically estimated marginal survival functions yield better precision and accuracy in the estimation results compared to the nonparametrically estimated marginal survival functions.
This observation confirms the additional uncertainty induced by the Beran's estimator. 
Table~\ref{MISE_Clayton} also indicates that the estimation performance improves with increasing sample size. 
Similar conclusions are reached in the simulations under the Frank and Gumbel copulas (see Tables \ref{MISE_Frank} and \ref{MISE_Gumbel}).  
A graphical representation of the results is provided in Figure \ref{fig:estimation_Clayton} for the convex model with sample size $n=250$ under the Clayton copula. 
We observe that, on average, the underlying functional form is estimated successfully, with slightly wider confidence intervals for nonparametrically estimated margins and at higher censoring rates.

\medskip

Next, we evaluate the proposed testing strategy through the empirical type I error and the empirical power attained at the significance level $\alpha = 0.05$. Under each of the considered scenarios, we perform the test focusing on the cases with sample size $n=250$ and considering $M=200$ replicates. For each replicate, $B=100$ bootstrap samples are drawn to obtain an approximate $p$-value. The rejection rates are reported in Table~\ref{test_Clayton}. 

\medskip

\begin{table}[h!]
\footnotesize
\caption{The empirical type I error (constant model) and the empirical power (convex and concave models) at significance level $5\%$ under the Clayton copula. }
\label{test_Clayton}
\centering
\vspace{0.1in}
\begin{tabular}{ccccccc}
\toprule \\ [-3.5ex]
&&&&& \multicolumn{2}{c}{} \\[-1.75ex]
&&&& Parametric Margins && Nonparametric Margins  \\[0.5ex]
\cmidrule(r){5-5} \cmidrule(r){7-7}
 & censoring rate & n && rejection rate && rejection rate\\
\midrule
\multirow{3}{*}{Constant}
& 0 \% & 250  && 0.080 && 0.085\\
& 20 \% & 250 && 0.085 && 0.150\\
& 50 \% & 250 && 0.090 && 0.170
\\ [0.8ex] \midrule
\multirow{3}{*}{Convex}
& 0 \% & 250  && 0.975 && 0.845\\
& 20 \% & 250 && 0.980 && 0.815\\
& 50 \%& 250  && 0.765 && 0.630
\\ [0.8ex] \midrule
\multirow{3}{*}{Concave}
& 0 \% & 250  && 0.990 && 0.605\\
& 20 \% & 250 && 0.990 && 0.500\\
& 50 \%& 250  && 0.865 && 0.305
 \\ [0.8ex]
\bottomrule
\end{tabular}
\end{table}

As can be seen in Table~\ref{test_Clayton}, the empirical type I error rates, obtained under the constant model, are higher than the nominal level $\alpha=0.05$, which indicates that the proposed test is liberal. Further, the empirical type I error tends to be higher when the conditional margins are estimated nonparametrically and when the censoring rate is higher. 
The rejection rates reported under the convex and concave models allow the assessment of the empirical power of the test.
Overall, we observe a higher empirical power for the parametrically estimated margins than for the nonparametrically estimated margins. The empirical power decreases as the censoring rate increases. The test results under the Frank and Gumbel copulas provide further evidence for these conclusions (see Tables \ref{test_Frank} and \ref{test_Gumbel}). 

A comparison of the results across different copula families indicates a better performance in the estimation and in the testing for the Frank and Gumbel copulas than for the Clayton copula. 
Note that the Clayton copula has lower tail dependence, while the Frank and Gumbel copulas exhibit no tail dependence and upper tail dependence, respectively. 
In the context of joint survival models, a copula with lower (upper) tail dependence represents the association between late (early) event times. 
Typically, late event times are subject to right-censoring; hence the loss of information occurs mainly in the lower tail area. 
It follows that the Clayton copula is affected more by the right-censoring, leading to a relatively inferior performance in the estimation and in the testing strategy.

\vspace*{-8pt}
\section{Data Analysis} 
\label{sec:5}

In this section, we analyze a subset of the data from the Diabetic Retinopathy Study, which was considered in \cite{Huster/Brookmeyer/Self:1989}, among others.
This study was conducted to investigate the effectiveness of laser photocoagulation in delaying the onset of blindness in diabetic retinopathy patients.
 One eye of each patient was randomly selected for treatment and the other eye was observed without treatment.
 The patients were followed up until their visual acuity got below a threshold at two consecutive visits.
The analysis subset consists of $n=197$ high risk patients, with censoring rates 73\% for the treated eyes and 49\% for the untreated eyes.
The data on the first and the last patients are given by $(Y_{11},Y_{21},\delta_{11},\delta_{21},X_1)=(46.23,46.23,0,0,28)$ and $(Y_{1n},Y_{2n},\delta_{1n},\delta_{2n},X_n)=(41.93,41.93,0,0,32)$, respectively. 
{}{Note that the event times are subject to univariate censoring.}
 More details on the dataset can be found in \cite{Huster/Brookmeyer/Self:1989}.

Besides assessing the effectiveness of the laser photocoagulation treatment, of interest was also to understand the dependence between the times to blindness of the treated and untreated eyes.
For our purposes, we consider the age at onset of the diabetic retinopathy as a continuous covariate and investigate if it has a significant impact on this dependence.

We fit the conditional joint survival function of the event times given the age at onset in two stages, as outlined in Section \ref{sec:2}.
Following \cite{Huster/Brookmeyer/Self:1989}, we use the Weibull model for the conditional marginal survival functions and the Clayton copula for the conditional dependence between the event times given the age at onset. 
The parameter estimates for the conditional marginal survival functions under the Weibull model are given in Table \ref{table:DRS Weibull}.
These estimates are based on the Weibull parametrization in \cite{Duchateau/Janssen:2008}, and are in close correspondence with the ones given in \cite{Huster/Brookmeyer/Self:1989}.
We also employ the Beran's estimator with the Epanechnikov kernel and obtain nonparametric estimates of each conditional margin at $10$ bandwidth values, ranging from $3$ to $57$ on a logarithmic scale.
The same $10$ bandwidth values are considered in the local likelihood estimation.
For the conditional copula model with parametrically estimated conditional margins, the optimal bandwidth value is $h_\mathbb{C} =42$.
When the conditional margins are estimated nonparametrically using the Beran's estimator, we performed bandwidth selection over a grid of $10 \times 10\times 10$ candidate values and obtained the optimal bandwidth vector $(h_1, h_2,h_\mathbb{C} ) = (3, 3, 42)$.

\begin{figure}[t!]
\center{
\includegraphics[width=0.95\textwidth]{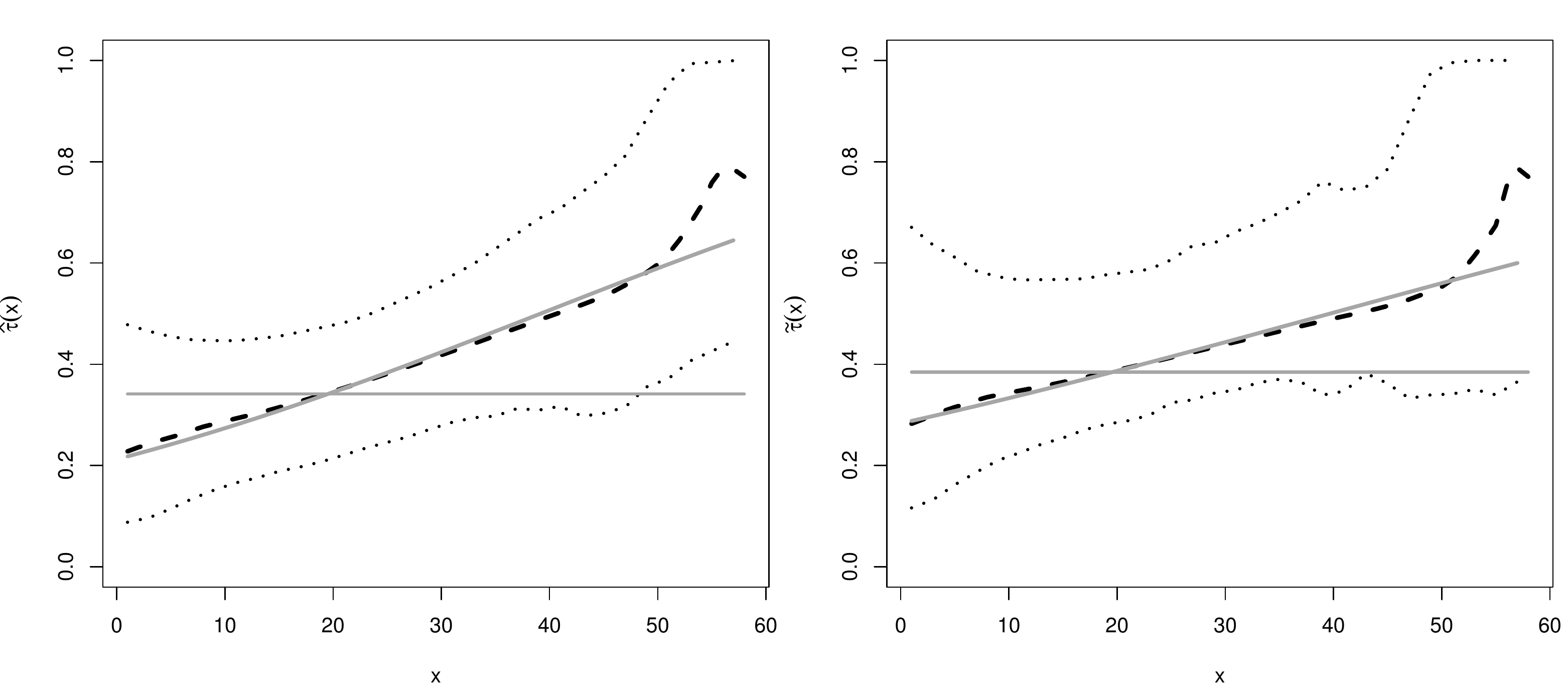}
\caption{Local likelihood estimates of the Kendall's tau (dashed lines) as a function of the age at onset of diabetic retinopathy obtained from parametrically (left panel) and nonparametrically (right panel) estimated conditional survival functions, along with the $90\%$ bootstrap confidence intervals (dotted lines) under the Clayton copula.  Also shown are the maximum likelihood estimates of the Kendall's tau (grey solid lines) obtained under the constant and linear calibration models.
\label{fig:drs2}}
}
\end{figure}

The local likelihood estimates of Kendall's tau at the selected bandwidths under each type of conditional margins are shown in Figure \ref{fig:drs2}, together with the $90\%$ bootstrap confidence intervals.
The latter are obtained by resampling the original data points $B=1000$ times, and fitting a joint model for each bootstrap sample at the bandwidth values selected for the original data.
The results from the parametric and nonparametric conditional margins both suggest an increasing linear pattern in the strength of dependence with the age at onset of diabetic retinopathy.
For comparisons, we also fit constant and linear calibration models using maximum likelihood. The resulting estimates are displayed in Figure \ref{fig:drs2} in the Kendall's tau scale.
As can be seen, the local likelihood estimates are in close agreement with the parametric estimates under the linear calibration model.
Furthermore, we observe slightly wider bootstrap confidence intervals, hence a larger variation in the Kendall's tau estimates when the Beran's estimator is used.
This observation is in line with our findings in Section \ref{sec:4}. 
There is more uncertainty in the local likelihood estimates when the age at onset is greater than 40. This is mainly due to the limited number of patients (31 out of 197) with high onset age, for which most of the observations are censored for at least one eye. 

To decide whether the observed variation in the strength of dependence (ranging approximately from $0.2$ to $0.7$ in the Kendall's tau scale) is significant or not, we perform the generalized likelihood ratio test as outlined in Section \ref{sec:3}.
The $p$-values based on $B=1000$ bootstrap samples under the null hypothesis are $0.138$ for the parametric conditional margins and $0.540$ for the nonparametric conditional margins.
Hence, there is not enough evidence in the data to reject the constant conditional copula model.
Note that the traditional likelihood ratio test between the constant and linear calibration models also supports this conclusion with $p$-values $0.111$ and $0.275$ for parametrically and nonparametrically estimated conditional margins, respectively.
These results together suggest that the impact of the age at onset on the dependence between the times-to-blindness in the treated and untreated eyes is not statistically significant.
Nevertheless, the observed increasing pattern in the strength of dependence may be of clinical interest, and would be worth further investigation in a larger sample.

\vspace*{-8pt}
\section{Discussion} 
\label{sec:6}

In this paper we outline an estimation and a testing strategy to assess the impact of a continuous cluster-level covariate on the strength of within-cluster dependence of right-censored event time data, as modelled via a conditional copula. 
A local likelihood approach is used to estimate the functional form of the conditional copula parameter and a generalized likelihood ratio test is described to test its constancy. 
A bootstrap procedure is employed to obtain an approximate $p$-value for the test.
The performance of the estimation and the testing method is evaluated in a simulation study, under different rates of right-censoring and for various parametric copula families, considering both parametrically and nonparametrically estimated margins. 
The results indicate that the local likelihood approach leads to on target estimation, with more uncertainty for nonparametrically estimated margins than for parametrically estimated margins, and that, depending on the considered parametric copula family, the testing strategy has reasonable to high power. 

The simulation results further suggest that the proposed bootstrap strategy is not optimal in terms of the accuracy of the test.
The high type 1 error rates may be as a result of the bandwidth parameter choice in the bootstrap samples. 
To reduce the computational burden, we use the bandwidth values for the original data, as selected by the cross-validated loglikelihood criterion, also in the bootstrap samples. 
Alternatively, one can perform bandwidth selection for each bootstrap sample, and approximate the null distribution of the test statistic with the corresponding bootstrap test values. 
This would, however, incur a rather high computational cost, and does not guarantee an accurate test decision.
A more feasible option is to perform the test at a specific (but not data-driven) bandwidth value for both the original data and the bootstrap samples. 
The higher the value of the bandwidth, the more conservative the test becomes. 
Our investigations for the simulated data suggest that when a global bandwidth value is used, the rejection rate of the test under the null model is much closer to the nominal type 1 error rate, yet still attains a reasonable power under the alternative models, especially when the conditional margins are estimated parametrically. 
As noted before, the presence of censoring and the nonparametric convergence rates of the conditional copula estimator as well as of the Beran's estimator, together make an assessment of the asymptotic null distribution quite cumbersome.
Nevertheless, the simulation study suggests that the use of a bootstrap is a valid alternative.

In the implementation of the proposed methods, it is assumed that the copula family is correctly specified. However, in practice, one has to select a copula family appropriate for the within-cluster dependence of the right-censored event time data being analyzed. This selection can be based on a likelihood-based criterion, such as Akaike information criterion as typically employed in copula modelling, or by comparing the predictive performance of competitive copula models as suggested in \cite{Acar/Craiu/Yao:2011}. In our analysis of the diabetic retinopathy data, we followed \cite{Huster/Brookmeyer/Self:1989} and employed the Clayton copula. The estimation results under the Frank and Gumbel copula yield a similar pattern for the impact of age at onset on the strength of dependence between the time to blindness of treated and untreated eyes (see Figures \ref{fig:drsF} and \ref{fig:drsG}). 

The procedures in this paper are developed for clustered right-censored event time data, assuming that the censoring times are independent of the event times and the covariate. 
This assumption is required to obtain the likelihood function in \eqref{eq3}, which contains only the distribution of the event times but not the censoring distribution.
Extending the methods to event time data subject to dependent censoring requires a careful consideration, due to the problem of non-identifiability of copula \citep{Wang:2012}.
On the other hand, extensions of the methods to other types of event time data with independent censoring are possible with some additional efforts, i.e. adjustments are needed in the estimation of the conditional margins as well as in the estimation of the conditional copula parameter. 
A particular future research direction is to develop conditional copula models for interval censored data.

\section*{Acknowledgements}

Funding in support of this work was provided by the Hasselt University Visiting Scholar Program (BOF Short Research Stay), the Natural Sciences and Engineering Research Council (NSERC) of Canada and the Canadian Statistical Sciences Institute (CANSSI). Financial support from the Interuniversity Attraction Poles Programme (IAP-network P7/06), Belgian Science Policy Office, is gratefully acknowledged. The computational resources and services used in this work were provided by the VSC (Flemish Supercomputer Center), funded by the Hercules Foundation and the Flemish Government – department EWI.

\bibliographystyle{asa}
\bibliography{GAJ-bibliography}

\clearpage
\vspace*{-8pt}
\section*{Appendix} 

\begin{algorithm}
{\footnotesize
\vspace{0.1cm}
\begin{algorithmic}
\State {\bf Step 1.} Obtain $\lambda_n (h_\mathbb{C})$ from the sample.
\State {\bf Step 2.} For $b=1,\ldots,B$ generate resamples in the following way:

\begin{itemize}
\setlength{\itemindent}{3.8em}
\item[{\bf Step 2.1.}] Generate $(U_{1i}^{b}, U_{2i}^{b})$ from the copula $\mathbb{C}$ with $\widehat{\theta}_0$ as the copula parameter value.
\item[{\bf Step 2.2.}] Obtain $(T_{1i}^{b}, T_{2i}^{b})$ via $T_{ki}^{b}=\widehat{S}_{k|x}^{-1}(U_{ki}^{b}|X_i)$, for $k=1,2$.
\item[{\bf Step 2.3.}] Generate $(C_{1i}^{b}, C_{2i}^{b})$ under one of the following scenarios:

\begin{itemize}
\setlength{\itemindent}{5.4em}
\item[{\bf Step 2.3a.}] {\bf Non-univariate censoring:} If $\delta_{ki}=0$, set $C_{ki}^{b}=Y_{ki}$. If $\delta_{ki}=1$, generate $C_{ki}^{b}$ from the conditional censoring distribution given that $C_{ki}>Y_{ki}$, i.e. generate $C_{ki}^{b}$ from
$$
    \frac{\widetilde{G}_k(t)-\widetilde{G}_k(Y_{ki})}{1-\widetilde{G}_k(Y_{ki})},
$$
where $\widetilde{G}_k$ is the Kaplan-Meier estimator of the censoring distribution based on the observations $(Y_{ki},1-\delta_{ki})$ for $k=1,2$ and $i = 1, \ldots,n$.

\item[{\bf Step 2.3b.}] {\bf Univariate censoring:} if $\delta_{ki}=0$ for at least one $k$ ($k=1,2$), set $C_{i}^{b}= \max(Y_{1i},Y_{2i})$; if $\delta_{ki}=1$ for all $k$ ($k=1,2$), generate $C_{i}^{b}$ from the conditional censoring distribution given that $C_i> Y_{i,\rm{max}} = \max\{Y_{1i},Y_{2i}\}$, i.e. generate $C_{i}^{b}$ from
$$
    \displaystyle \frac{\widetilde{G}_k(t)-\widetilde{G}_k(Y_{i,\rm{max}})}{1-\widetilde{G}_k(Y_{i,\rm{max}})},
$$
where $\widetilde{G}_k$ is the Kaplan-Meier estimator of the censoring distribution based on the observations $(Y_{i,\max},1-\delta_{1i}\delta_{2i})$, for $i = 1, \ldots,n$.
\end{itemize}

\item[{\bf Step 2.4.}] Set $Y_{ki}^{b}=\min\{T_{ki}^{b},C_{ki}^{b}\}$ and $\delta_{ki}^{b}=I(T_{ki}^{b} \leq C_{ki}^{b})$.
\item[{\bf Step 2.5.}]Set $\widehat{U}_{ki}^{b}=\widehat{S}_{k|x}^{b}(Y_{ki}^{b}|X_i)$ with $\widehat{S}_{k|x}^{b}$ the estimate of $S_{k|x}$ based on the observations $(Y_{ki}^{b}, \delta_{ki}^{b})$, for $k=1,2$ and $i = 1, \ldots,n$.
\item[{\bf Step 2.6.}] Fit the copula $\mathbb{C}$ to $(\widehat{U}_{1i}^{b}, \widehat{U}_{2i}^{b})$ and obtain the bootstrap value of the generalized likelihood ratio statistic: $\lambda_n^{b} (h_\mathbb{C})$.
\end{itemize}

\State {\bf Step 3.} Calculate the approximate p-value via
$$
p_{\rm{boot}}=\sum_{b=1}^B I\left( \lambda_n^{b} (h_\mathbb{C}) \geq \lambda_n (h_\mathbb{C})\right) /B.
$$
\end{algorithmic}
\caption{\small Bootstrap algorithm for the GLR statistic under parametrically estimated conditional marginal survival functions.}
\label{alg:1}
}
\end{algorithm}

\begin{algorithm}
{\footnotesize
\vspace{0.1cm}
\begin{algorithmic}

\State {\bf Step 1.} Obtain $\lambda_n (h_1, h_2, h_\mathbb{C})$ from the sample.
\State {\bf Step 2.} For $b=1,\ldots,B$ generate resamples in the following way:

\begin{itemize}
\setlength{\itemindent}{3.8em}
\item[{\bf Step 2.1.}] Generate $(U_{1i}^{b}, U_{2i}^{b})$ from the copula $\mathbb{C}$ with $\widetilde{\theta}_0$ as the copula parameter value.
\item[{\bf Step 2.2.}] Obtain $(T_{1i}^{b}, T_{2i}^{b})$ via $T_{ki}^{b}=\widetilde{S}_{k|x}^{-1}(U_{ki}^{b}|X_i)$, for $k=1,2$.
\item[{\bf Step 2.3.}] Generate $(C_{1i}^{b}, C_{2i}^{b})$ under one of the following scenarios:

\begin{itemize}
\setlength{\itemindent}{5.4em}
\item[{\bf Step 2.3a.}] {\bf Non-univariate censoring:} If $\delta_{ki}=0$, set $C_{ki}^{b}=Y_{ki}$. If $\delta_{ki}=1$, generate $C_{ki}^{b}$ from the conditional censoring distribution given that $C_{ki}>Y_{ki}$, i.e. generate $C_{ki}^{b}$ from
$$
    \frac{\widetilde{G}_k(t)-\widetilde{G}_k(Y_{ki})}{1-\widetilde{G}_k(Y_{ki})},
$$
where $\widetilde{G}_k$ is the Kaplan-Meier estimator of the censoring distribution based on the observations $(Y_{ki},1-\delta_{ki})$ for $k=1,2$ and $i = 1, \ldots,n$.

\item[{\bf Step 2.3b.}] {\bf Univariate censoring:} if $\delta_{ki}=0$ for at least one $k$ ($k=1,2$), set $C_{i}^{b}= \max(Y_{1i},Y_{2i})$; if $\delta_{ki}=1$ for all $k$ ($k=1,2$), generate $C_{i}^{b}$ from the conditional censoring distribution given that $C_i> Y_{i,\rm{max}} = \max\{Y_{1i},Y_{2i}\}$, i.e. generate $C_{i}^{b}$ from
$$
    \displaystyle \frac{\widetilde{G}_k(t)-\widetilde{G}_k(Y_{i,\rm{max}})}{1-\widetilde{G}_k(Y_{i,\rm{max}})},
$$
    where $\widetilde{G}_k$ is the Kaplan-Meier estimator of the censoring distribution based on the observations $(Y_{i,\max},1-\delta_{1i}\delta_{2i})$, for $i = 1, \ldots,n$.
\end{itemize}

\item[{\bf Step 2.4.}] Set $Y_{ki}^{b}=\min\{T_{ki}^{b},C_{ki}^{b}\}$ and $\delta_{ki}^{b}=I(T_{ki}^{b} \leq C_{ki}^{b})$.
\item[{\bf Step 2.5.}]Set $\widetilde{U}_{ki}^{b}=\widetilde{S}_{k|x}^{b}(Y_{ki}^{b}|X_i)$ with $\widetilde{S}_{k|x}^{b}$ the estimate of $S_{k|x}$ based on the observations $(Y_{ki}^{b}, \delta_{ki}^{b})$, for $k=1,2$ and $i = 1, \ldots,n$.
\item[{\bf Step 2.6.}] Fit the copula $\mathbb{C}$ to $(\widetilde{U}_{1i}^{b}, \widetilde{U}_{2i}^{b})$ and obtain the bootstrap value of the generalized likelihood ratio statistic: $\lambda_n^{b} (h_1,h_2,h_\mathbb{C})$.
\end{itemize}

\State {\bf Step 3.}  Calculate the approximate p-value via 
$$
p_{\rm{boot}}=\sum_{b=1}^B I\left( \lambda_n^{b} (h_1, h_2, h_\mathbb{C}) \geq \lambda_n (h_1, h_2, h_\mathbb{C}) \right) /B.
$$\end{algorithmic}
\caption{Bootstrap algorithm for the GLR statistic under nonparametrically estimated conditional marginal survival functions.}
\label{alg:2}
}\end{algorithm}

\clearpage
 
\setcounter{table}{0}
\renewcommand{\thetable}{S\arabic{table}}%
\setcounter{figure}{0}
\renewcommand{\thefigure}{S\arabic{figure}}

\section*{Supplemental Material} 

 \bigskip

{\bf  \large Simulation Results under the Frank and Gumbel copulas}

\bigskip

\begin{table}[h!] \footnotesize
\caption{Integrated Squared Bias, Integrated Variance and Integrated Mean Square Error (multiplied by 100) of the Kendall's tau estimates under the Frank copula.}
\label{MISE_Frank}
\centering
\vspace{0.1in}
\begin{tabular}{cccccccccccccc}
\toprule \\ [-1.5ex]
&&&&& \multicolumn{6}{c}{} &&& \\[-1.75ex]
&&&&  \multicolumn{3}{c}{Parametric Margins}&&&\multicolumn{3}{c}{Nonparametric Margins}  \\[0.5ex] \cmidrule(r){5-8} \cmidrule(r){10-13}
& $\%$ cens & n && IBIAS$^{2}$ & IVAR & IMSE &&& IBIAS$^{2}$ & IVAR & IMSE \\
\midrule
\multirow{6}{*}{Constant}
& 0 & 250  && 0.003 & 0.544 & 0.547 &&& 0.014 & 0.639 & 0.653\\
& & 500    && 0.002 & 0.264 & 0.266 &&& 0.005 & 0.307 & 0.311 \\
& 20 & 250 && 0.002 & 0.657 & 0.659 &&& 0.022 & 0.792 & 0.813 \\
& & 500    && 0.001 & 0.329 & 0.330 &&& 0.010 & 0.375 & 0.385 \\
& 50 & 250 && 0.005 & 1.069 & 1.074 &&& 0.031 & 1.162 & 1.194 \\
& & 500    && 0.002 & 0.501 & 0.503 &&& 0.010 & 0.579 & 0.589
\\ [0.8ex] \midrule
\multirow{6}{*}{Convex}
& 0 & 250  && 0.173 & 1.556 & 1.729 &&& 0.258 & 1.587 & 1.845\\
& & 500    && 0.098 & 0.756 & 0.854 &&& 0.144 & 0.795 & 0.938 \\
& 20 & 250 && 0.180 & 1.857 & 2.037 &&& 0.261 & 1.937 & 2.198 \\
& & 500    && 0.097 & 0.932 & 1.029 &&& 0.150 & 0.959 & 1.109 \\
& 50 & 250 && 0.308 & 2.853 & 3.161 &&& 0.404 & 2.934 & 3.339 \\
& & 500    && 0.152 & 1.556 & 1.708 &&& 0.201 & 1.527 & 1.728
\\ [0.8ex] \midrule
\multirow{6}{*}{Concave}
& 0 & 250  && 0.078 & 1.045 & 1.123 &&& 0.104 & 1.224 & 1.328\\
& & 500    && 0.034 & 0.562 & 0.596 &&& 0.037 & 0.603 & 0.640 \\
& 20 & 250 && 0.093 & 1.122 & 1.215 &&& 0.115 & 1.349 & 1.464 \\
& & 500    && 0.048 & 0.615 & 0.663 &&& 0.049 & 0.661 & 0.710 \\
& 50 & 250 && 0.142 & 1.683 & 1.825 &&& 0.214 & 1.898 & 2.111 \\
& & 500    && 0.073 & 0.920 & 0.993 &&& 0.086 & 1.009 & 1.095
 \\ [0.8ex]
\bottomrule
\end{tabular}
\end{table}

\begin{table}[h!] \footnotesize
\caption{Integrated Squared Bias, Integrated Variance and Integrated Mean Square Error (multiplied by 100) of the Kendall's tau estimates under the Gumbel copula.}
\label{MISE_Gumbel}
\centering
\vspace{0.1in}
\begin{tabular}{cccccccccccccc}
\toprule \\ [-1.5ex]
&&&&& \multicolumn{6}{c}{} &&& \\[-1.75ex]
&&&&  \multicolumn{3}{c}{Parametric Margins}&&&\multicolumn{3}{c}{Nonparametric Margins}  \\[0.5ex] \cmidrule(r){5-8} \cmidrule(r){10-13}
& $\%$ cens & n && IBIAS$^{2}$ & IVAR & IMSE &&& IBIAS$^{2}$ & IVAR & IMSE \\
\midrule
\multirow{6}{*}{Constant}
& 0 & 250  && 0.005 & 0.581 & 0.586 &&& 0.015 & 0.802 & 0.817\\
& & 500    && 0.001 & 0.286 & 0.287 &&& 0.008 & 0.370 & 0.378 \\
& 20 & 250 && 0.006 & 0.690 & 0.696 &&& 0.010 & 0.973 & 0.974 \\
& & 500    && 0.001 & 0.334 & 0.336 &&& 0.004 & 0.440 & 0.444 \\
& 50 & 250 && 0.006 & 1.176 & 1.182 &&& 0.022 & 1.420 & 1.442 \\
& & 500    && 0.003 & 0.504 & 0.507 &&& 0.005 & 0.651 & 0.656
\\ [0.8ex] \midrule
\multirow{6}{*}{Convex}
& 0 & 250  && 0.089 & 1.688 & 1.777 &&& 0.239 & 1.946 & 2.184 \\
& & 500    && 0.068 & 0.812 & 0.880 &&& 0.181 & 0.853 & 1.035 \\
& 20 & 250 && 0.102 & 2.570 & 2.672 &&& 0.202 & 2.889 & 3.092 \\
& & 500    && 0.074 & 0.981 & 1.056 &&& 0.181 & 1.055 & 1.236 \\
& 50 & 250 && 0.169 & 3.847 & 4.016 &&& 0.328 & 3.740 & 4.068 \\
& & 500    && 0.086 & 1.497 & 1.583 &&& 0.236 & 1.585 & 1.820
\\ [0.8ex] \midrule
\multirow{6}{*}{Concave}
& 0 & 250  && 0.108 & 0.988 & 1.096 &&& 0.121 & 1.219 & 1.341\\
& & 500    && 0.051 & 0.506 & 0.557 &&& 0.046 & 0.620 & 0.666 \\
& 20 & 250 && 0.125 & 1.109 & 1.234 &&& 0.123 & 1.368 & 1.491 \\
& & 500    && 0.056 & 0.549 & 0.605 &&& 0.054 & 0.674 & 0.728 \\
& 50 & 250 && 0.145 & 1.591 & 1.736 &&& 0.166 & 1.895 & 2.062 \\
& & 500    && 0.087 & 0.837 & 0.924 &&& 0.084 & 0.964 & 1.048
 \\ [0.8ex]
\bottomrule
\end{tabular}
\end{table}

\clearpage

\begin{table}[h!] \footnotesize
\caption{The empirical type I error (constant model) and the empirical power (convex and concave models) at significance level $5\%$ under the Frank copula. }
\label{test_Frank}
\centering
\vspace{0.1in}
\begin{tabular}{ccccccc}
\toprule \\ [-1.5ex]
&&&&& \multicolumn{2}{c}{} \\[-1.75ex]
&&&& Parametric Margins && Nonparametric Margins  \\[0.5ex] \cmidrule(r){1-7}
 & censoring rate & n && rejection rate && rejection rate\\
\midrule
\multirow{3}{*}{Constant}
& 0 \% & 250  && 0.110 && 0.105\\
& 20 \% & 250 && 0.085 && 0.085\\
& 50 \% & 250 && 0.095 && 0.100
\\ [0.8ex] \midrule
\multirow{3}{*}{Convex}
& 0 \% & 250  && 0.960 && 0.935\\
& 20 \% & 250 && 0.945 && 0.915\\
& 50 \%& 250  && 0.805 && 0.745
\\ [0.8ex] \midrule
\multirow{3}{*}{Concave}
& 0 \% & 250  && 0.985 && 0.975\\
& 20 \% & 250 && 0.970 && 0.920\\
& 50 \%& 250  && 0.850 && 0.735
 \\ [0.8ex]
\bottomrule
\end{tabular}
\end{table}

\bigskip
\bigskip
\bigskip
\bigskip

\begin{table}[h!] \footnotesize
\caption{The empirical type I error (constant model) and the empirical power (convex and concave models) at significance level $5\%$ under the Gumbel copula. }
\label{test_Gumbel}
\centering
\vspace{0.1in}
\begin{tabular}{ccccccc}
\toprule \\ [-1.5ex]
&&&&& \multicolumn{2}{c}{} \\[-1.75ex]
&&&& Parametric Margins && Nonparametric Margins  \\[0.5ex] \cmidrule(r){1-7}
 & censoring rate & n && rejection rate && rejection rate\\
\midrule
\multirow{3}{*}{Constant}
& 0 \% & 250  && 0.090 && 0.090\\
& 20 \% & 250 && 0.080 && 0.090\\
& 50 \% & 250 && 0.115 && 0.125
\\ [0.8ex] \midrule
\multirow{3}{*}{Convex}
& 0 \% & 250  && 0.980 && 0.875\\
& 20 \% & 250 && 0.950 && 0.845\\
& 50 \%& 250  && 0.835 && 0.730
\\ [0.8ex] \midrule
\multirow{3}{*}{Concave}
& 0 \% & 250  && 0.985 && 0.935\\
& 20 \% & 250 && 0.990 && 0.875\\
& 50 \%& 250  && 0.880 && 0.755
 \\ [0.8ex]
\bottomrule
\end{tabular}
\end{table}

\clearpage
{\bf \large Details on the Analysis of Diabetic Retinopathy Data}

\begin{table}[h!] \footnotesize
\caption{Parameter estimates and their standard errors (in parentheses) for the conditional marginal survival functions of the event times given the age at onset  under the Weibull model.}
\label{table:DRS Weibull}
\centering
\vspace{0.1in}
\begin{tabular}{ccc cc ccc}
\toprule \\ [-2ex]
 \multicolumn{3}{c}{Treated Eye} && \multicolumn{3}{c}{Untreated Eye}  \\[0.5ex]
$\rho_1$ & $\lambda_1$ & $\beta_1$   &&   $\rho_2$ & $\lambda_2$ & $\beta_2$   \\
\midrule
0.788 & 0.021 & -0.015 && 0.830 & 0.022 & 0.014 \\
(0.099) & (0.009)& (0.010) &&  (0.074) & (0.007)& (0.007) \\
\bottomrule
\end{tabular}
\end{table}

\bigskip
\bigskip
\bigskip
\bigskip

\begin{figure}[h!]
\center{
\includegraphics[width=0.95\textwidth]{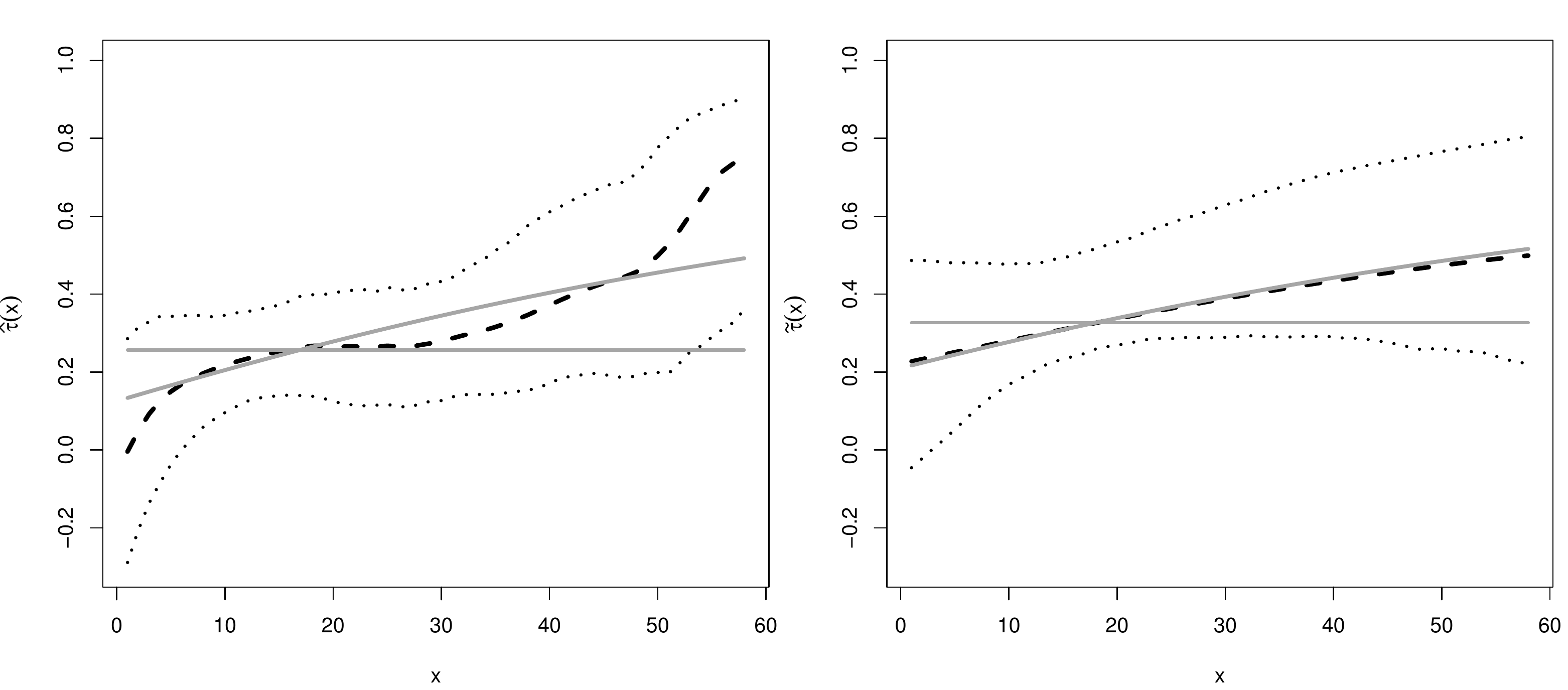}
\caption{Local likelihood estimates of the Kendall's tau (dashed lines) as a function of the age at onset of diabetic retinopathy obtained from parametrically (left panel) and nonparametrically (right panel) estimated conditional survival functions, along with the $90\%$ bootstrap confidence intervals (dotted lines) under the Frank copula. Also shown are the maximum likelihood estimates of the Kendall's tau (grey solid lines) obtained under the constant and linear calibration models.
\label{fig:drsF}}
}
\end{figure}

\begin{figure}[h!]
\center{
\includegraphics[width=0.95\textwidth]{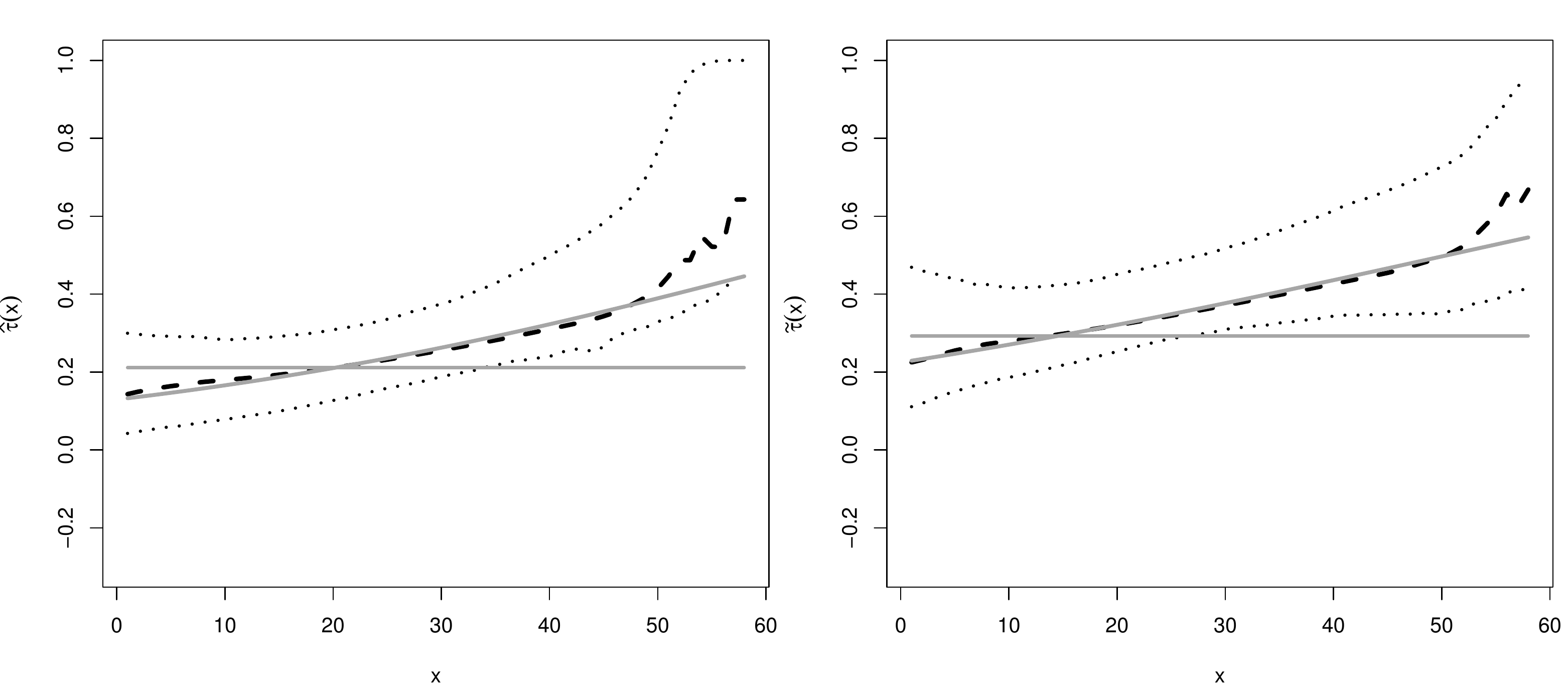}
\caption{Local likelihood estimates of the Kendall's tau (dashed lines) as a function of the age at onset of diabetic retinopathy obtained from parametrically (left panel) and nonparametrically (right panel) estimated conditional survival functions, along with the $90\%$ bootstrap confidence intervals (dotted lines) under the Gumbel copula. Also shown are the maximum likelihood estimates of the Kendall's tau (grey solid lines) obtained under the constant and linear calibration models.
\label{fig:drsG}}
}
\end{figure}

\bigskip

\begin{table}[h!] \footnotesize
\caption{Test results for the impact of the age at onset of diabetic retinopathy on the dependence between the times-to-blindness in the treated and untreated eyes under the Clayton, Frank and Gumbel copulas. Provided below are the bootstrap $p$-values for the generalized likelihood ratio test and the approximate $p$-values for the traditional likelihood ratio test when the conditional survival functions are estimated parametrically (left panel) and nonparametrically (right panel). Under each case, the selected bandwidth values are also reported. }
\label{table:DRS Test}
\centering
\vspace{0.1in}
\begin{tabular}{c ccc cc ccc}
\toprule \\ [-2ex]
 & \multicolumn{3}{c}{Parametric Margins} && \multicolumn{3}{c}{Nonparametric Margins}  \\[1ex]
& GLRT   & LRT & $h_\mathbb{C} $      &&   GLRT   & LRT & $(h_1, h_2,h_\mathbb{C} )$    \\[2ex]
\midrule
Clayton & 0.138 & 0.111 & 42 && 0.540 & 0.275 & (3, 3, 42) \\[1ex]
Frank & 0.125  & 0.120 & 23  && 0.440 & 0.221 & (3, 3, 57) \\[1ex]
Gumbel & 0.290 & 0.148& 42 && 0.515 &0.200& (5, 3, 42)\\
\bottomrule
\end{tabular}
\end{table}

\end{document}